# Classification of diffraction patterns in single particle imaging experiments performed at X-ray free-electron lasers using a convolutional neural network

Alexandr Ignatenko[1], Dameli Assalauova[1], Sergey A. Bobkov[2], Luca Gelisio[1], Anton B. Teslyuk[2,3], Viacheslav A. Ilyin[2,3] and Ivan A. Vartanyants[1,4,*]

[1]*Deutsches Elektronen-Synchrotron DESY, Notkestrasse 85, Hamburg, 22607 Germany*
[2]*National Research Centre "Kurchatov Institute", pl. Akademika Kurchatova 1, Moscow, 123182 Russia*
[3]*Moscow Institute of Physics and Technology, Institutskiy per. 9, Dolgoprudny, Moscow region, 141701 Russia*
[4]*National Research Nuclear University MEPhI, Kashirshkoe sh. 31, 115409 Moscow, Russia*

**August 17, 2020**

**Abstract**

Single particle imaging (SPI) is a promising method for native structure determination which has undergone a fast progress with the development of X-ray Free-Electron Lasers. Large amounts of data are collected during SPI experiments, driving the need for automated data analysis. The necessary data analysis pipeline has a number of steps including binary object classification (single versus multiple hits). Classification and object detection are areas where deep neural networks currently outperform other approaches. In this work, we use the fast object detector networks YOLOv2 and YOLOv3. By exploiting transfer learning, a moderate amount of data is sufficient for training of the neural network. We demonstrate here that a convolutional neural network (CNN) can be successfully used to classify data from SPI experiments. We compare the results of classification for the two different networks, with different depth and architecture, by applying them to the same SPI data with different data representation. The best results are obtained for YOLOv2 color images linear scale classification, which shows an accuracy of about 97% with the precision and recall of about 52% and 61%, respectively, which is in comparison to manual data classification.

[*]Corresponding author: Ivan.Vartaniants@desy.de

**Introduction**

Single Particle Imaging (SPI) enables structural investigations of non-crystalline objects [1, 2]. Many particles with random orientation are illuminated by an X-ray free-electron laser (XFEL) beam, which consists of millijoule pulses with durations of tens of femtoseconds [3, 4, 5]. Scattering patterns of the investigated particles are therefore collected before radiation damage takes place, in a so-called 'diffraction before destruction' [6] experiment. Since SPI does not require crystallization, it enables investigation of objects which are hard – or impossible – to crystallize, like viruses, proteins, organelles, and others [7, 8, 9, 10].

A general analysis pipeline for SPI data consists of several steps as first proposed in [2] and further extended in [11, 12]. One of the important steps is the classification of all diffraction patterns measured in a XFEL experiment, and specifically identification of single hits, i.e. events containing the scattering pattern of a single particle. Earlier this task was performed by different methods such as: principle component analysis (PCA) or support vector machine (SVM) [13]. In a recent work [12], classification of single hits was performed using expectation-maximization (EM) algorithms developed in cryogenic electron microscopy [14]. In this work, we propose to select single hits from experimental data set using a convolutional neural network (CNN) approach. CNNs are known to give excellent results for the task of image classification and object detection [15, 16]. CNN-based solutions have been successfully applied recently to classify diffraction patterns collected in coherent diffraction imaging experiments at XFELs [17,18] and tomography experiments at synchrotron sources [19].

Here, we present the results of classification of diffraction patterns in SPI experiments using a CNN. We use transfer learning to train the network for classification with the limited amount of training and validation data, and demonstrate that this approach is comparable with the manual selection method.

**2. Methods**

*2.1 Experiment and data preparation*

The SPI experiment was performed at the Atomic Molecular Optics (AMO) instrument [20, 21] at the Linac Coherent Light Source (LCLS) at SLAC National Accelerator Laboratory in the frame of the SPI initiative [22] (experiment AMOX34117 [23]). Samples of PR772 bacteriophage [23, 24] were aerosolized using a gas dynamic virtual nozzle (GDVN) in a helium environment [25]. The particles were injected into the sample chamber using an aerodynamic lens injector [8, 26]. The samples in the particle stream intersected the focused and pulsed XFEL beam. The XFEL had a repetition rate of 120 Hz, an average pulse energy of ~2 mJ, a focus size of ~1.5 μm, and a photon energy of 1.7 keV (wavelength 0.729 nm). Diffraction patterns were recorded by the pnCCD detector [28] mounted at 0.130 m distance from the interaction region. The size of the panel was 512 by 1024 pixels with a pixel size of 75×75 μm$^2$.

The total amount of experimental data collected was of the order of $1.2 \times 10^9$ patterns [23]. Only part of the data that were classified as hits contained scattering signal in the diffraction pattern. The hit finding was performed using the software "psoake" in the "psana" framework [28]. As a result, about $1.9 \times 10^7$ diffraction patterns were selected as hits from the initial set of experimental data [23]. A mask was introduced to remove bad pixels, areas of the detector with high background and saturation. The background scattering signal near the center of the diffraction pattern was subtracted and particle size filtering was performed with a particle size range from 55 nm to 84 nm being selected as described in [12]. As an outcome of this size filtering, 18,213 patterns were selected for single hit classification. In this work a single hit is defined as a diffraction pattern corresponding to X-ray scattering on a single particle and is distinguished from a multiple hit that originates from X-ray scattering on few particles or scattering from other objects (for example, water droplets). The goal of this work is to perform binary classification of measured data and to sort diffraction patterns belonging to a "single hit" class from other patterns.

The selection of single hits by the CNN was compared with the manual selection performed by the visual inspection of the patterns [23]. The number of single hits for the manual selection method was 1,393. From this selection we considered only the sub-set of 1,196 single hits that belongs to a data set filtered by a particle size in the range from 55 nm to 84 nm. Visual inspection of the manual selection did not reveal any false positive patterns and for this reason we considered the manual selection as a ground truth.

*2.2 CNN choice*

The choice of a CNN for our task was based on the following consideration: binary classification of single hits for the SPI experiments is different from the classification of features for the diffraction patterns in the experiments described in [18]. Indeed, binary classification is easier than having multiple cases. On the other hand, diffraction patterns in SPI experiments contain scattering signal from particles in many arbitrary orientations. It is therefore more difficult for a CNN to learn that all diffraction patterns contain an image of the same particle, and to distinguish them from all other possible kind of diffraction patterns.

In this work we used a fast object detector YOLOv2 [29] from the open source neural network framework Darknet [30] to perform classification. Its relatively shallow CNN is similar to the one used in Ref. [18]. We compare the performance of YOLOv2 with the more recent and deeper version YOLOv3 [31]. This deeper network was considered in order to understand, whether a significant gain in performance may be obtained between these two networks. This could also justify an increase in complexity of calculations by using YOLOv3 network. Both YOLO-versions are real-time object detection systems, they run faster than other detection methods with comparable performance.

*2.3 CNN description*

CNNs consist of layers, the core building blocks being convolutional layers. Together with other layers, such as pooling layers, they form a network [32]. The layers are characterized by a set of filters or kernels, containing weights. The weights are adjusted during training of the CNN. The set of weights for the specific CNN at a certain training stage forms a model with the weights being its parameters. Additionally, there are parameters that are adjusted beyond the training process. They are called hyper-parameters. Examples of hyper-parameters are the number and size of the filters in the layers and the learning rate.

The CNN algorithm is a supervised deep learning algorithm. It infers a function that maps an input and output from the training data which consists of a set of training examples with annotations. They represent a desired output that is referred to as a ground truth. These annotations are class labels in the case of classification. For the object detectors they are supplemented by the position of the object [33, 34], typically represented by a bounding box surrounding the object. The neural networks are trained using a gradient descent optimization algorithm [35]. The optimization algorithm minimizes the so-called loss function by updating the parameters of the model. The parameters are updated according to the chain rules [36]. The learning rate is the hyper-parameter that controls how much the parameters of the model are changed in response to the model error. CNNs usually use a stochastic gradient descent (SGD) algorithm [37, 38] for optimizing the loss function during training. The weights are updated based on random subsets (batches) of the training data rather than the complete training set. The period during training when the network has seen one batch is called iteration.

The YOLOv2 has a plain architecture with a limited depth (see Figure 1(a)). The base of YOLOv2 is the Darknet-19 classification network consisting of 19 convolutional layers. It was pre-trained on 1,000 classes of images from the ImageNet [39]. The YOLOv2 program accepts images of any size, converts them to the size of 416×416 pixels, and uses them as an input for the CNN. The YOLOv3 software (see Figure 1(b)), contrary to YOLOv2, has a residual network architecture [40] and is based on the Darknet-53 classification network with 53 convolutional layers. The residual blocks are necessary to avoid possible performance degradation with the growing depth of the network. Darknet-53 was pre-trained on the same 1,000 classes of the data set at ImageNet. YOLOv3 converts input images to the size of 608×608 pixels and uses them as an input for the CNN.

The loss functions of YOLOv2 and YOLOv3 as object detectors have terms associated with classification and localization (see Refs. [29] and [31] for details). We have modified the detection part of the networks to perform detection of objects of one class (single hits). An image was classified as a single hit when the network detected a single hit in the image. In the opposite case, when the network did not detect a single hit in the image, we classified it as not a single hit (binary classification). Default detection thresholds of 0.24 and 0.5 were applied for the networks YOLOv2 and YOLOv3, respectively.

*2.4 Metrics*

The main metrics to evaluate the classification results in the case of binary classification are accuracy, precision, and recall. Accuracy is a ratio of correct predictions among the total number of predictions made. Accuracy alone may be not informative if the frequency of single hits is a small fraction of the whole data set and it is therefore also useful to know precision and recall. Precision is a measure of classifier exactness. Recall is a measure of classifier completeness. These metrics are defined as

$$accuracy = \frac{TP + TN}{TP + TN + FP + FN}, \qquad (1)$$

$$\text{precision} = \frac{TP}{TP + FP}, \qquad (2)$$

$$\text{recall} = \frac{TP}{TP + FN}, \qquad (3)$$

where TP and TN are the number of true positive and true negative results and FP and FN are the number of false positives and false negatives results.

The F-scores metrics convey the balance between precision and recall. If the same importance is given to precision and recall, *i.e.* false positives are as undesirable as false negatives and then the $F_1$-score can be used

$$F_1 = 2\frac{precision * recall}{precision + recall}. \tag{4}$$

The metrics to compare the different outcome of classification is the intersection over union (IoU). ). In our case it is defined as

$$\text{IoU} = \frac{Inersection}{Union} = \frac{Inersection}{S_1 + S_2 - Intersection}, \tag{5}$$

where *Intersection* is the number of patterns classified as single hits that are common for the given CNN model and a reference method, $S_1$ is the number of patterns classified as single hits by the CNN model, $S_2$ is the number of patterns classified as single hits by the manual selection method.

*2.5 Image processing*

Below we describe an image processing step for conversion of the experimental data to be used as input to the CNN. The central part of each diffraction pattern with the size of 123x240 detector pixels containing the major part of the photon counts was selected. Then, it was up-sampled to the size of 954x1855 pixels and saved as an image using either a color or grayscale scheme, and either linear or logarithmic scale. This conversion scheme with the initial up-sampling ensures that each detector pixel is represented as a monochrome rectangle in the image used as an input to the neural network. A trained CNN extracts features like edges and corners from the sharp borders between the monochrome rectangles representing the detector pixels. Starting and processed diffraction patterns to be used as input for YOLOv2 are shown in Figure 2. For the color representation, photon counts either in linear or logarithmic scale were converted into three RGB layer images according to "jet" color map. For the grayscale representation, photon counts in linear and logarithmic scale were converted into one grayscale layer. In our implementation the three identical grayscale layers were stacked into a three RGB layer image. With such conversion all data were normalized to the maximum value of intensity of each diffraction pattern. Three color channels for linear and logarithmic representation of color images used as an input to the CNN are presented in Figure 3.

*2.6 Training and validation*

Building a data set to train the model is a crucial step. A good training set should be representative and balanced. That means that the training set should represent the data that the model attempts to describe as good as possible. The total number of training examples should be sufficiently large to train the CNN for the full depth. Our aim is to use transfer learning with a limited amount of training data. To achieve this goal, we utilized pre-trained weights, *i.e.* the weights that were obtained by training with a large data set (ImageNet in our case) as a starting point for our specific training set with a limited amount of data. As a result of CNN training the weights in the last layers are mostly affected and adjusted to the selected small training data set.

The ratio of the number of examples for each class should be close to what is expected in the experimental data set. However, it is difficult to achieve this goal using a small training data set. In our case, based on our previous experience, we expect that the number of positive examples (single hit) will be much lower than the number of negative ones (no single hit). If we would consider similar ratio of positive and negative examples in the training set we could have a situation that negative examples will be selected with high probability during each iteration. As a compromise, our training set contained 165 positive and 390 negative examples taken from the experimental set of 18,213 patterns.

A validation set is used to evaluate a given model, the same requirements as for the training set apply to it as well. We kept the total number of examples small, although the ratio of positive to negative examples could be lower in the case of validation. The validation set in our case contained 53 positive and 283 negative examples taken, again, from the same set of experimental data.

The images for color (linear and logarithmic scale) and grayscale (linear and logarithmic scale) representations showing positive and negative examples of our training set are displayed in Figure 4. All training data had ground truth annotations. Annotations for the positive examples contain class labels (single hit) and the coordinates of the bounding box covering the major part of all intensity counts in the image. To accelerate the annotation process, we considered the same coordinates of the bounding box for all positive examples in the test set of the data as shown in Figure 4(a, b, e, f, i, j, m, n). The validation data set had annotations in the form of the class labels only (single hit or no single hit).

During training the value of the loss function (training loss) was calculated after each iteration (see Figure 5(a, c, e, g) and Figure 6(a, c)). One can see that the training loss was gradually going down as a function of iterations, which was a good sign of CNN convergence. The local increase of the training loss in some plots corresponds to the point, where the training rate was changed. The set of weights for the CNN was saved after 100 iterations, giving a model as was described above. A set of models at different training stages for the certain network architecture (YOLOv2 or YOLOv3) and data representation (color or grayscale images, linear or logarithmic scale) constitute a family of models. We make binary classification on the validation set applying the YOLO software with the weights obtained during the training process which is performed for each family of models. The results of classification at validation step were compared to the ground truth and the $F_1$-score was calculated (see Figure 5(b, d, f, h) and Figure 6(b, d)). The $F_1$-score goes to saturation after about 2,000, iterations for every family of models. It means that every model in the family may be regarded as optimal after 2,000 iterations. Table 1 summarizes the results of the mean values and population standard deviation of accuracy, precision, recall, and $F_1$-score calculated for the validation set. As it follows from this Table, the model families have high value of accuracy and precision, but significantly lower values of recall. This behavior is due to the fact that in validation set we have a large number of negative examples which are correctly identified. This means that in this case we have a comparably large number of TN that determines high values of accuracy. At the same time, we have a small number of FP cases that effectively makes precision value high as well. In the case of recall its value is determined by comparably large value of FN, which is on the same order as TP. This is due to the fact that CNN does not recognize all diffraction patterns labelled as positive in the validation set.

3. **Results**

We used NVIDIA GeForce GTX 1060 with 6 GB of internal memory for the training, validation and final tests using our CNN. The training time for 100 iterations was on average about 280 s for YOLOv2 and 290 s for YOLOv3. This training time was almost the same for both networks due to smaller batch size for the deeper network YOLOv3. The processing time for one image was about 20 ms for YOLOv2 and about 75 ms for YOLOv3. The difference in processing time between YOLOv2 and YOLOv3 was due to the difference in image size and depth of the network.

We performed classification on the data set of 18,213 diffraction patterns. We considered 4,000 iterations as optimal as soon as it was far from the saturation value of 2,000 iterations. As soon as $F_1$-score was showing comparably large population standard deviation we considered five consecutive models for each family (different by 100 iterations starting from the 4,000-th iteration). The number of single hits selected by each of these models for each network architecture and different data representation is given in Table 2. First of all, we observed how stable each model was. The population standard deviation in the number of selected single hits at different training stages for a given model was taken as a measure of stability. These stages are considered to have almost the same and optimal level of training and should give nearly the same amount of single hits, if the model is stable. As seen in Table 2, the spread in the number of selected single hits among the models of the same architecture (YOLOv2 or YOLOv3) is due to relatively small size of the training set, arbitrary orientation, and finite number of positive examples in the training set. The YOLOv2 models have significantly narrower distribution in the number of selected single hits contrary to the YOLOv3 models. Less stability for the YOLOv3 models can be explained by the smaller batch size and larger number of parameters due to the higher depth of the network.

In order to mitigate the instabilities in single hit selection we applied the following strategy. The diffraction pattern was classified as a single hit only if it was classified as a single hit in each of the five consequent models (see Table 2). We see from this Table that the lowest standard deviation is observed for YOLOv2 color images case. In fact, this has happened occasionally due to a small distribution of $F_1$-score values in this region of iterations (see Figure 5(b)).

In order to compare performance of different networks we evaluated intersection and IoU of our results with the manual selection set of data. Inspection of Table 3 shows that performance of both networks for color images in linear scale is similar. The performance of YOLOv2 model trained on grayscale images in linear scale is significantly better than that of YOLOv3. This serves as an indication that relatively shallow neural network is sufficient for single hit classification. The models trained on grayscale images perform worse than that trained on color images, which is more pronounced in the case of YOLOv3. Features extracted from three different color layers seem to have more information than the ones extracted from the one grayscale layer. We expected that the model with logarithmic scale of intensities should perform significantly better than the linear scale one as soon as the scattering signal decays as $I(q) \sim q^{-3} \div q^{-4}$ [41]. Table 3 shows that both models perform rather similar. We think that this can be due to the fact that the features extracted by the CNN from color representation of diffraction patterns in linear or logarithmic scale are similar. This may be valid for the range of low total intensities that we have in the considered SPI experiment. We also noted (see Table 3), that the grayscale logarithmic images perform similar to color images, however, performance of the grayscale linear images is worse. We think that the feature extraction from the grayscale linear images is less informative in comparison to color images.

As it follows from our discussion good performance was observed practically for all the models except the ones trained on gray linear scale images (see Table 3). For example, for YOLOv2 color linear model our selection has 792 common selected single hits with the manual selection method and 525 common selected single hits with the EM-based selection method [12]. That provides IoU value of 44% with the manual selection method.

As discussed earlier, our training set consisting of 240 patterns was taken from the set of 18,213 experimental patterns. In order to avoid data evaluation bias, we removed these 240 training examples from the initial set of 18,213 patterns that gave us the test set of 17,973 patterns. This set was used for evaluation of accuracy, precision and recall. If we consider the manual selection as a ground truth, the accuracy for the YOLOv2 color linear model calculated for the test set of 17,973 patterns is around 97% with the precision and recall being 52% and 61%, respectively. Such result may be explained by the following considerations. Large value of accuracy is due to a large number of TN values. In this case, due to Eq. (1), accuracy may be close to 100%. This also indicates that accuracy alone is not sufficient metric to describe results of CNN classification. At the same time precision has become much lower than in the case of validation set due to a high value of FP (that is on the same order as TP) determined by the CNN, though visually part of them should belong to positive values. We notice also that recall values are similar to the ones determined on the validation set of data.

4. **Conclusions and Outlook**

In conclusion, classification of single hits in SPI experiments can be effectively performed by a CNN. We demonstrated that it is possible to extract single hits with high level of accuracy with respect to the manual selection. A moderate depth of the CNN, like the one of YOLOv2, was sufficient for this work and the use of a deeper network did not improved the efficiency. The spatial integrity of diffraction patterns seems not to be crucial for the work of CNN. We found a similar performance for the classification utilizing a CNN and the one performed by the EM-based method. We found also that efficiency of the CNN, on the data set that was used for our tests, does not depend on color linear or logarithmic scale. However, performance of the CNN was worse in the case of grayscale images. This question will need more detailed studies in the future.

We used transfer learning to train the network for classification with a limited amount of training data taken from experiment. Such a small training set can produce an additional noise in the selection of single hits (as seen in Table 2). To overcome this problem instead of SGD different variant of gradient descent method may be used. As an alternative approach training of the CNN on a large set of simulated data may be used.

Using a CNN with transfer learning, in our opinion, is crucial as this enables automatization of the analysis pipeline. One can think of having a person who identifies some positive and negative examples in the first few hours of experiment, train the network and let it run for the rest of the experiment. By that, one can even solve one of the most challenging throughput problems of today's experiments, *i.e.* data reduction in XFEL experiments. Indeed, one could even foresee that the network decides which data are stored and which are not saved at all.

An alternative approach would be to generalize the CNN model to be applicable for heterogeneous systems, when different conformations of a particle are studied in an experiment. In such case a convolutional auto-encoder (CAE) [42] can be used. The CAE consists of two parts named encoder and decoder trained together. The encoder is trying to learn a low-dimensional representation of the input data, while the decoder reconstructs the data using its low-dimensional representation. At the stage of training, when the input patterns will coincide with the output patterns, we may assume that the network is fully trained and this set of data may be used as a ground truth. The CAE can be trained from input to output using a large amount of simulated data for different conformation of particles used in the SPI experiments. When the training is finished, the encoder part trained to produce a low-dimensional representation of input data can be used as a base for the classification network.

The proposed approach based on CNN for classification of large amount of data may be beneficial for applications at high repetition rate XFELs [5] while collecting data with the Megahertz rate [43].

**Funding**

This work was supported by the Helmholtz Associations Initiative and Networking Fund and Russian Science Foundation (Grant No. HRSF-0002/18-41-06001).

**Acknowledgements**

We acknowledge the support of the project and discussions with E. Weckert, X. Yang for the careful reading of the manuscript, and A. Aquila for fruitful discussions in the frame of this project. The Helmholtz Association Innovation Pool Project AMALEA is acknowledged.

**References**

[1] Neutze R et al. 2000 Potential for biomolecular imaging with femtosecond X-ray pulses *Nature* **406** 752

[2] Gaffney J K and Chapman H N 2007 Imaging atomic structure and dynamics with ultrafast X-ray scattering *Science* **316** 1444

[3] Emma P et al. 2020 First lasing and operation of an ångstrom-wavelength free-electron laser *Nat. Photon.* **4** 641

[4] Ishikawa T et al. 2012 2012 A compact X-ray free-electron laser emitting in the sub-ångström region *Nat. Photon.* **6** 540

[5] Decking W et al. 2020 A MHz-repetition-rate hard X-ray free-electron laser driven by a superconducting linear accelerator *Nat. Photon.* **14**, 391

[6] Chapman H N et al. 2006 Femtosecond diffractive imaging with a soft-X-ray free-electron laser *Nat. Phys.* **2** 839

[7] Seibert M M et al. 2011 Single mimivirus particles intercepted and imaged with an X-ray laser *Nature* **470** 78

[8] Hantke M F et al. 2014 High-throughput imaging of heterogeneous cell organelles with an X-ray laser *Nat. Photon.* **8** 943

[9] van der Schot G et al. 2015 Imaging single cells in a beam of live cyanobacteria with an X-ray laser *Nat. Commun.* **6** 5704

[10] Ekeberg T et al. 2015 Three-dimensional reconstruction of the giant mimivirus particle with an X-ray Free-electron laser *Phys. Rev. Lett.* **114** 098102

[11] Rose M et al. 2018 Single-particle imaging without symmetry constraints at an X-ray free-electron laser *IUCrJ* **5** 727

[12] Assalauova D et al. 2020 An advanced workflow for single particle imaging with the limited data at an X-ray free-electron laser *IUCrJ* (submitted)

[13] Bobkov S A et al. 2015 Sorting algorithms for single-particle imaging experiments at X-ray free-electron lasers *J. Synchrotron Rad.* **22**, 1345

[14] Dempster A P, Laird N M and Rubin D B 1977 Maximum likelihood from incomplete data via the EM algorithm *J. R. Stat. Soc. Ser. B.* **39** 1

[15] Krizhevsky A, Sutskever I and Hinton G E 2012 ImageNet classification with deep convolutional neural networks *Proc. of Conf. Advances in Neural Information Processing Systems (NIPS)* 25

[16] Szegedy Ch, Toshev A and Erhan D 2013 Deep neural networks for object detection *Proc. of Conf. Advances in Neural Information Processing Systems (NIPS)* 2553

[17] Shi Y et al. 2019 Evaluation of the performance of classification algorithms for XFEL single-particle imaging data *IUCrJ* **6** 331

[18] Zimmerman J et al. 2019 Deep neural networks for classifying complex features in diffraction images *Phys. Rev. E* **99** 063309

[19] Yang X et al. 2020 Tomographic reconstruction with a generative adversarial network *J. Synchrotron Rad.* **27**, 486

[20] Ferguson K R et al. 2015 The atomic, molecular and optical science instrument at the Linac Coherent Light Source *J. Synchrotron Rad.* **22** 492

[21] Osipov T et al. 2018 The LAMP instrument at the Linac Coherent Light Source free-electron laser *Rev. Sci. Instrum.* **89** 035112

[22] Aquila A et al. 2015 The linac coherent light source single particle imaging road map *Struct. Dyn.* **2** 041701

[23] Li H et al. 2020 Diffraction data from aerosolized coliphage PR772 virus particles imaged with the Linac Coherent Light Source *Sci. Data* (submitted)

[24] Reddy H K N et al. 2017 Coherent soft X-ray diffraction imaging of coliphage PR772 at the Linac coherent light source *Sci. Data* **4** 170079

[25] Nazari R et al 3D printing of gas-dynamic virtual nozzles and optical characterization of high-speed microjets *Opt. Express* **28(15)** 21749

[26] Benner W H et al. 2008 Non-destructive characterization and alignment of aerodynamically focused particle beams using single particle charge detection *J. Aerosol Sci.* **39** 917

[27] Strüder L et al. 2010 Large-format, high-speed, X-ray pnCCDs combined with electron and ion imaging spectrometers in a multipurpose chamber for experiments at 4th generation light sources *Nucl. Instruments Methods Phys. Res. Sect. A Accel. Spectrometers, Detect. Assoc. Equip.* **614**, 483

[28] Damiani D et al. 2016 Linac Coherent Light Source data analysis using psana *J. Appl. Crystallogr.* **49** 672

[29] Redmon J and Farhadi A 2016 YOLO9000: Better, Faster, Stronger *arXiv:1612.08242*

[30] Darknet: Open Source Neural Networks in C https://pjreddie.com/darknet/

[31] Redmon J and Farhadi A 2018 YOLOv3: an incremental improvement *arXiv:1804.02767*

[32] Goodfellow I, Bengio Y and Courville A 2016 Deep Learning (*MIT Press*)

[33] Everingham M 2010 The Pascal visual object classes (VOC) challenge *Int. J. Comput. Vis.* **88** 303

[34] Russakovsky O et al. 2015 ImageNet Large Scale Visual Recognition Challenge *arXiv:*1409.0575

[35] Ruder S 2017 An overview of multi-task learning in deep neural networks *arXiv:*1706.05098

[36] Kaplan W, 1984 Advanced Calculus, 3rd ed (Addison-Wesley)

[37] Bottou L, Curtis F E and Nocedal J 2017 Layer normalization *arXiv:1607.06450*

[38] Goyal P et al. 2017 Accurate, large minibatch SGD: training ImageNet in 1 hour *arXiv:1706.02677*

[39] ImageNet http://www.image-net.org

[40] He K et al. 2015 Identity Mappings in Deep Residual Networks *arXiv:1603.05027*

[41] Rose M et al. 2018 Quantitative ptychographic bio-imaging in the water window *Opt. Express* **26(2)** 1237

[42] Masci J et al. 2011 Stacked Convolutional Auto-Encoders for Hierarchical Feature Extraction In: Honkela T., Duch W., Girolami M., Kaski S. (eds) Artificial Neural Networks and

Machine Learning – ICANN 2011. ICANN 2011. Lecture Notes in Computer Science, vol 6791. Springer, Berlin, Heidelberg

[43] Sobolev E et al. 2020 Megahertz single-particle imaging at the European XFEL *Commun. Phys.* **3** 97

**Table 1. Mean and population standard deviation (std) values (from 2000 to 5000 iterations) of the accuracy, precision, recall and $F_1$-score for the different families of YOLOv2 and YOLOv3 models calculated with respect to validation data set.**

| Family of models | Acc. mean, % | Acc. std, % | Prec. mean, % | Prec. std, % | Recall, mean, % | Recall, std, % | $F_1$-score mean, % | $F_1$-score std, % |
|---|---|---|---|---|---|---|---|---|
| **YOLOv2, color, linear** | 95 | 1 | 98 | 2 | 72 | 8 | 83 | 5 |
| **YOLOv2, color, log** | 94 | 1 | 92 | 4 | 64 | 9 | 76 | 7 |
| **YOLOv2, grayscale, linear** | 93 | 1 | 91 | 4 | 64 | 9 | 75 | 6 |
| **YOLOv2, grayscale, log** | 94 | 1 | 98 | 3 | 66 | 7 | 79 | 5 |
| **YOLOv3, color, linear** | 93 | 2 | 94 | 5 | 60 | 13 | 72 | 9 |
| **YOLOv3, grayscale, linear** | 89 | 1 | 83 | 7 | 39 | 8 | 53 | 6 |

**Table 2. Number of single hits classified by the models at different training stages, population standard deviation (std), and number of single hits in the stable selection.**

| Training stage<br>Model | N of iterations 4,000 | N of iterations 4,100 | N of iterations 4,200 | N of iterations 4,300 | N of iterations 4,400 | Population std | Number of single hits |
|---|---|---|---|---|---|---|---|
| **YOLOv2 color** | 1441 | 1572 | 1830 | 1804 | 1707 | 146.3 | 1379 |
| **YOLOv2 color log** | 1860 | 1830 | 1608 | 1989 | 2263 | 214.9 | 1561 |
| **YOLOv2 grayscale** | 3188 | 3246 | 2884 | 4053 | 2387 | 543.9 | 1945 |
| **YOLOv2 grayscale log** | 1702 | 2458 | 1799 | 1490 | 1111 | 441.5 | 1089 |
| **YOLOv3 color** | 1256 | 3707 | 4781 | 3245 | 2162 | 1222.4 | 1234 |
| **YOLOv3 grayscale** | 3641 | 3563 | 3319 | 2723 | 5208 | 824.1 | 2498 |

**Table 3**. Comparison of the stable selection for different CNN architecture, and data representation with the manual selection, which is considered as a ground truth. This manuall selection is chosen to estimate the Intersection, IoU, accuracy, precision, and recall.

| Model | Number of single hits | Intersection with the manual selection | IoU for manual selection, % | Accuracy, % | Precision, % | Recall, % |
|---|---|---|---|---|---|---|
| **YOLOv2 color, linear** | 1379 | 792 | 44 | 97 | 52 | 61 |
| **YOLOv2 color log** | 1561 | 808 | 42 | 97 | 46 | 62 |
| **YOLOv2 grayscale, linear** | 1945 | 813 | 35 | 97 | 36 | 63 |
| **YOLOv2 grayscale log** | 1089 | 676 | 42 | 97 | 55 | 50 |
| **YOLOv3 color, linear** | 1234 | 701 | 41 | 97 | 50 | 52 |
| **YOLOv3 grayscale, linear** | 2498 | 647 | 21 | 96 | 21 | 47 |

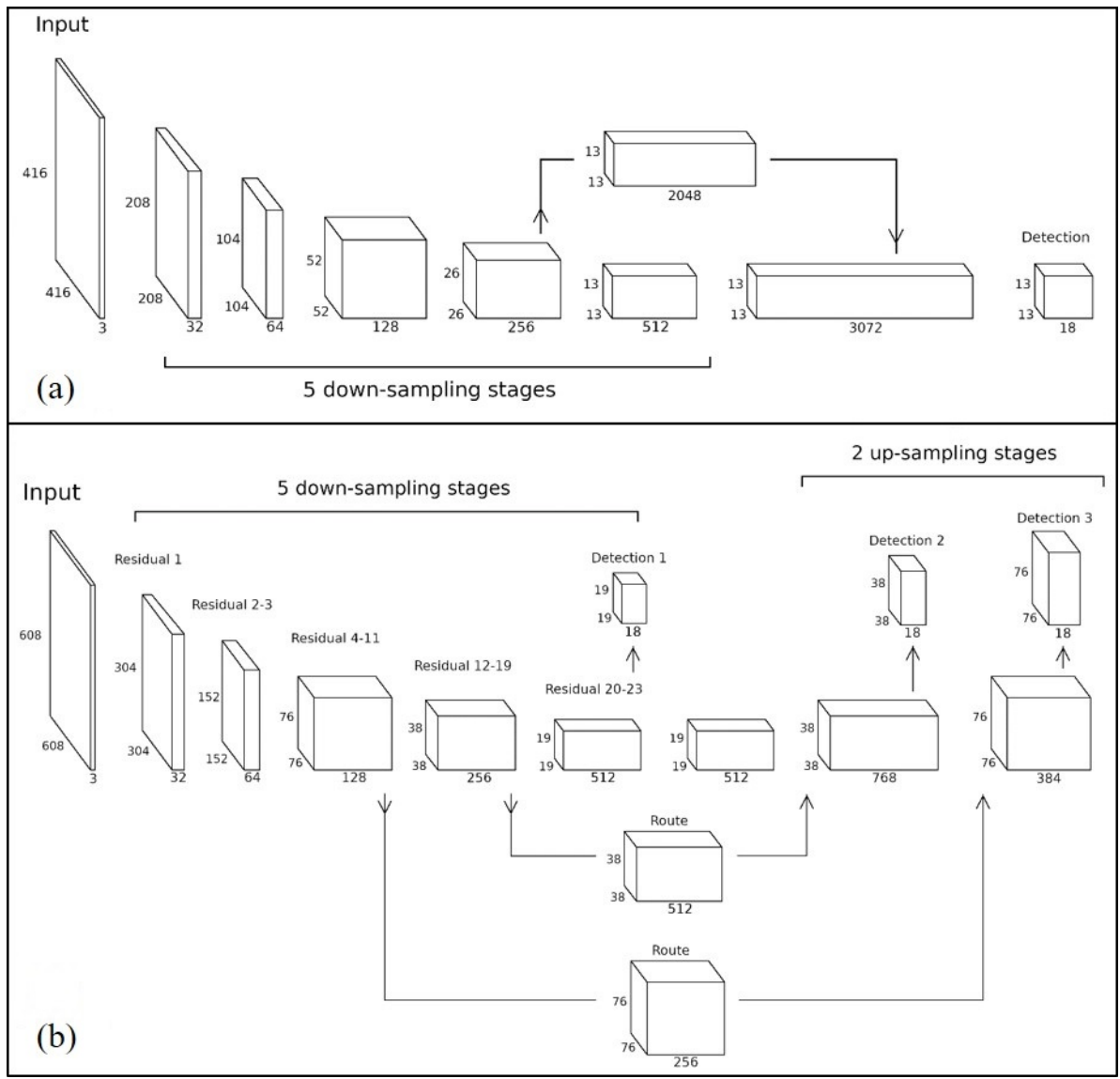

**Figure 1.** Architecture of CNN realized in YOLOv2 (a) and YOLOv3 (b). Both networks accept images with three color layers and a fixed size of 416×416 pixels (YOLOv2) and 608×608 pixels (YOLOv3). The data undergoes five down-sampling stages to limit the number of parameters. The dimensions of the input for every stage are shown in the figure. In case of YOLOv2 the data flows consequently from one layer to another. In case of YOLOv3 five down-sampling stages are followed by two up-sampling stages. Every up-sampling stage receives additional activation from the corresponding down-sampling stage (marked "Route"). Every down-sampling stage contains one to several residual blocks (marked "Residual").

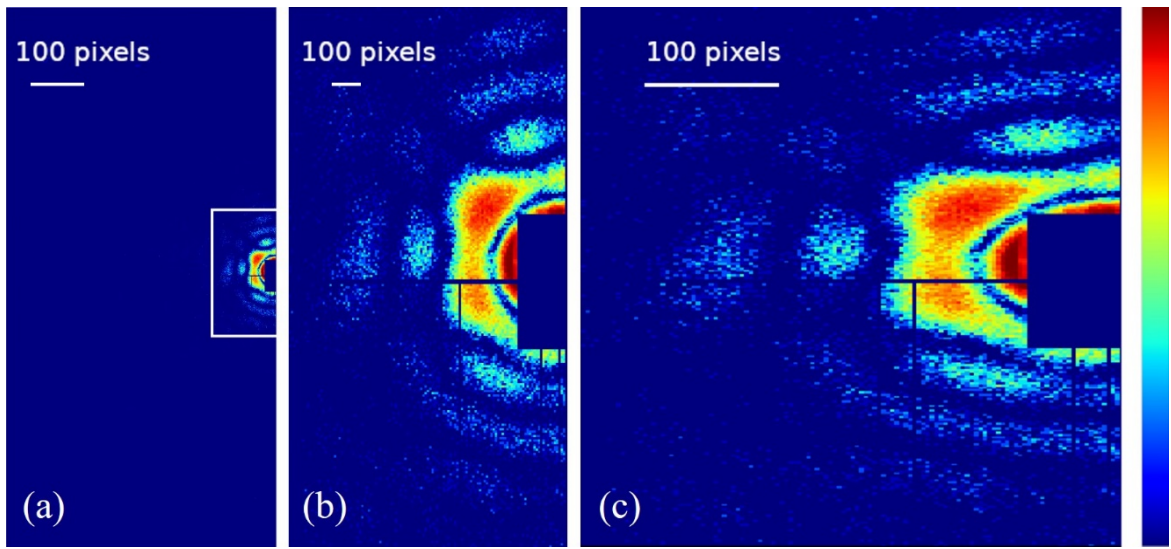

**Figure 2.** (a) Half of detector panel used in the experiment, white rectangle around the center of diffraction pattern shows the area that was used for the conversion; (b) the central part of diffraction pattern converted into color image and used for the analysis, (c) image Type equation here.modified by YOLOv2 ready to be used as an input to the neural network. Intensity in all three images is shown in logarithmic scale.

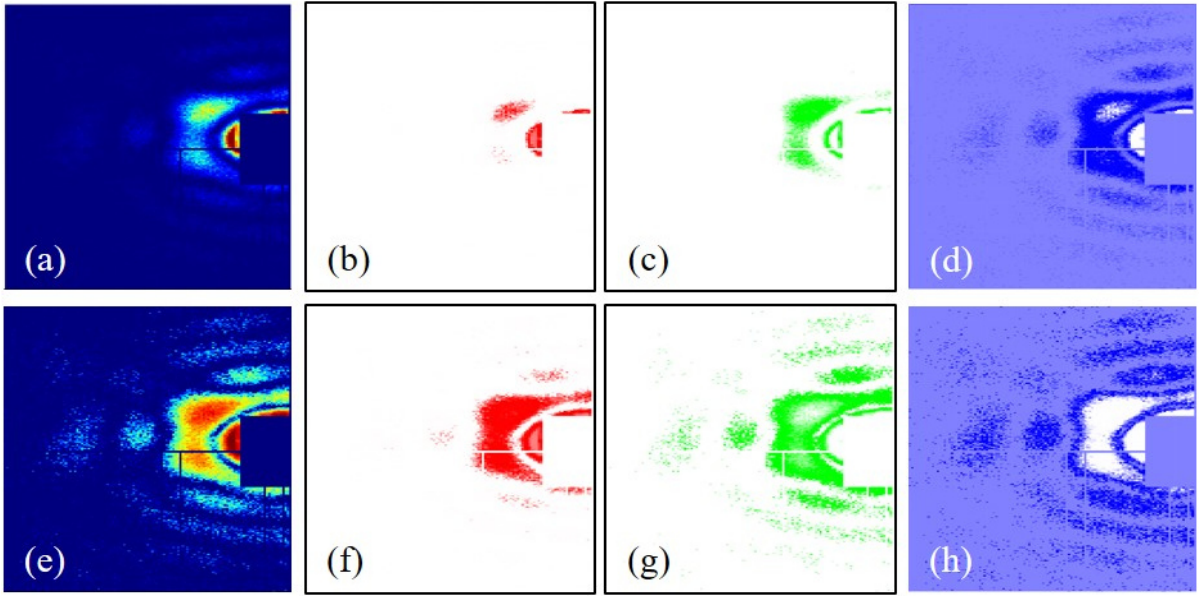

**Figure 3.** Constituent color layers for the two examples of the color images used for the CNN input linear (a) and logarithmic (e) scale (red (b, f), green (c, g) and blue (d, h)).

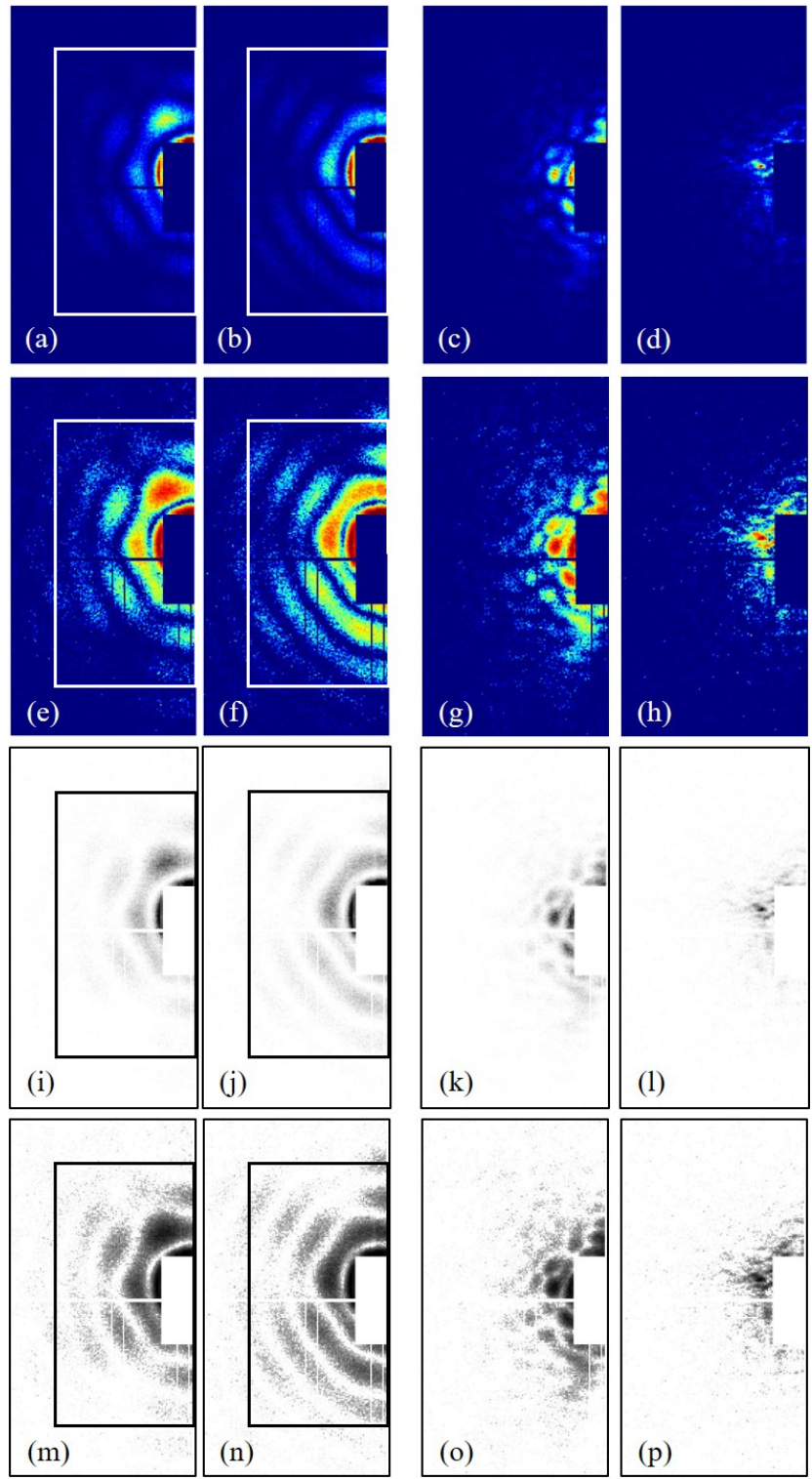

**Figure 4.** Different representations of the training data: 3 RGB layers, "jet" color scheme, linear scale to photon counts (a)-(d); 3 RGB layers, "jet" color scheme, logarithmic scale to photon counts (e)-(h); 3 identical grayscale layers, linear scale to photon counts (i)-(l); 3 identical grayscale layers, logarithmic scale to photon counts (m)-(p) Positive examples (single hits) – (a), (b), (e), (f), (i), (j), (m) and (n); negative examples – (c), (d), (g), (h), (k), (l), (o) and (p). Corresponding bounding box annotation is shown for positive examples.

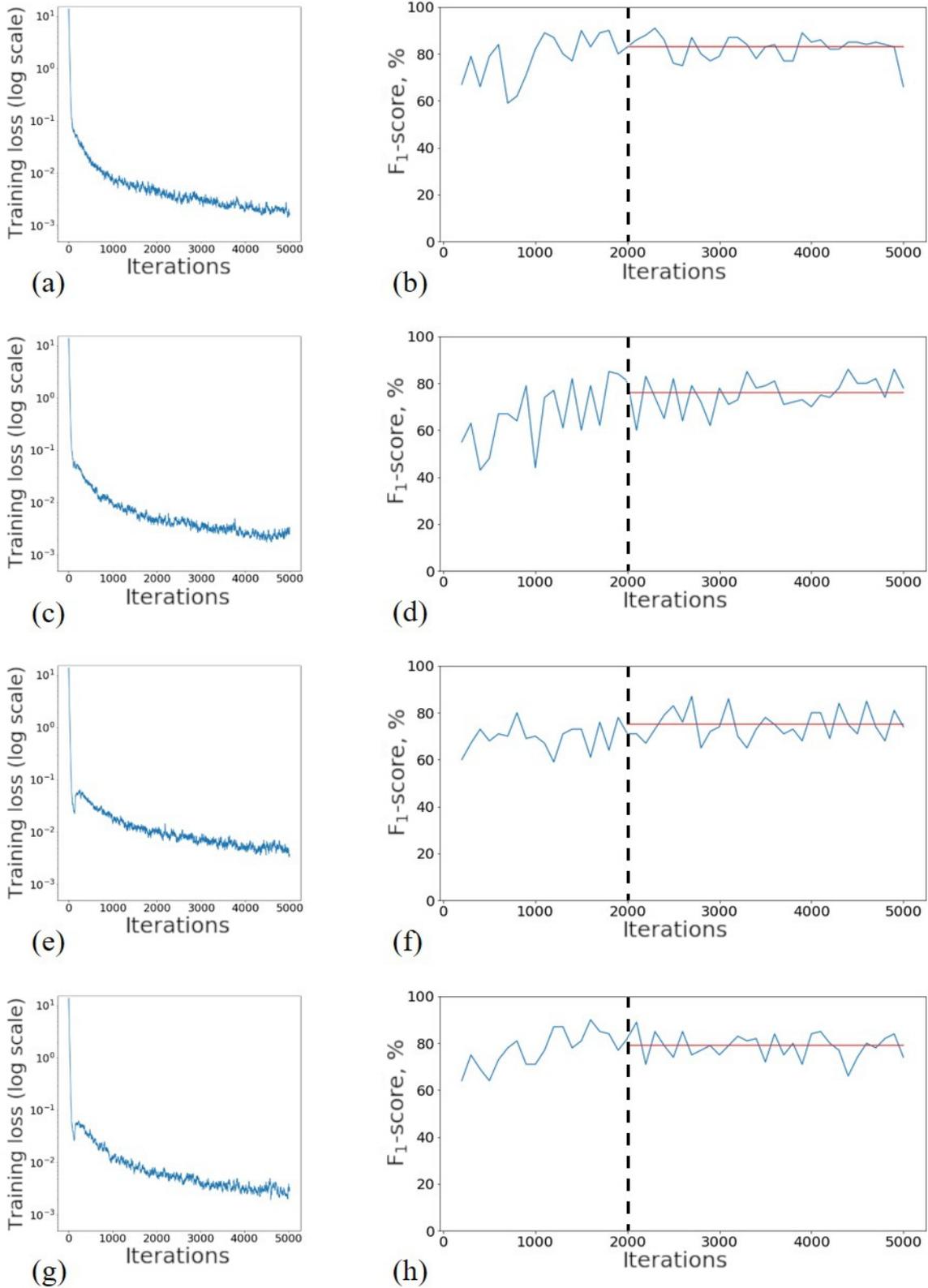

**Figure 5.** Training loss (left column) and $F_1$-score (right column) as a function of iteration calculated for validation data set and shown for the different model families of YOLOv2: (a, b) color image, linear scale, (c, d) color image, logarithmic scale, (e, f) grayscale image, linear scale, (g, h) grayscale image, logarithmic scale. Saturation of $F_1$-score is reached after 2000 iterations (shown by black vertical dashed line). Red lines show mean value of $F_1$-score at saturation level.

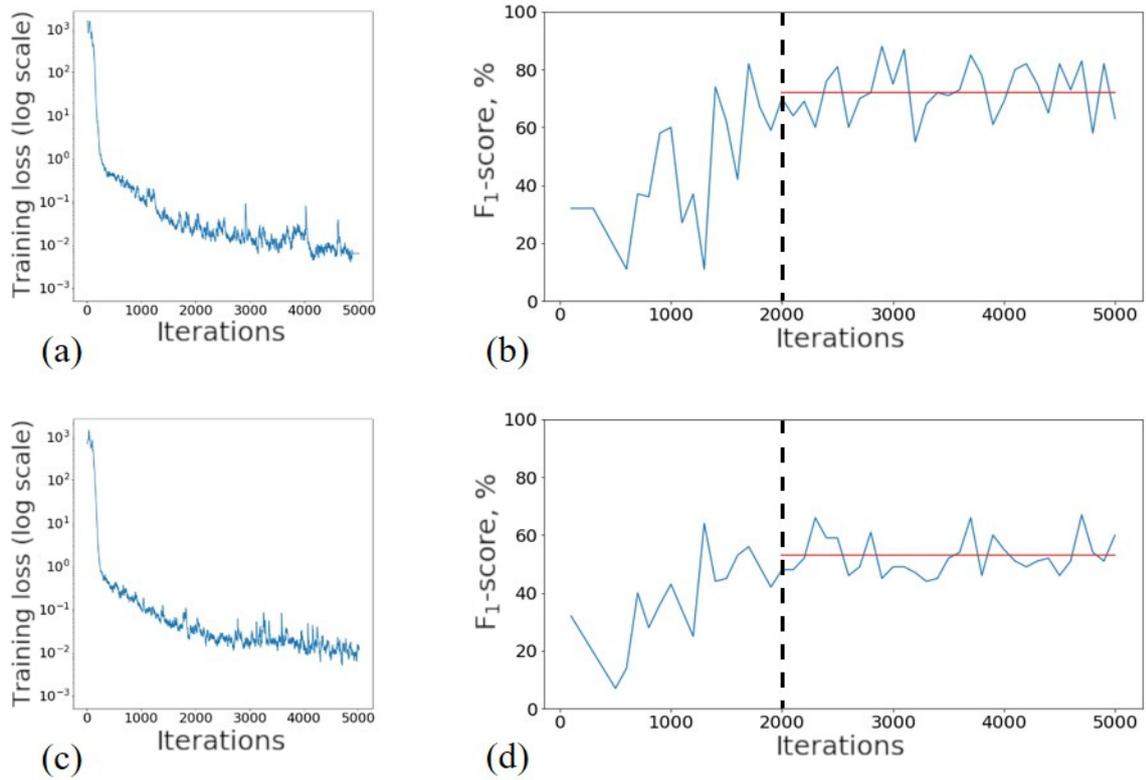

**Figure 6.** The same as in Fig. 6 for the different model families of YOLOv3: (a, b) color image, linear scale, (c, d) grayscale image, linear scale.